\documentstyle[epsf,12pt]{article}
\def\bn{\bigskip\noindent}
\begin{document}
\begin{titlepage}
\renewcommand{\thefootnote}{\fnsymbol{footnote}}
 \font\csc=cmcsc10 scaled\magstep1
 {\baselineskip=14pt
 \rightline{
 \vbox{\hbox{hep-th/0001037}
       \hbox{UT-872}
       }}}
\bn\bn\bn\bn\bn
\begin{center}
\Large{\bf{Boundary states in Gepner models}}
\vspace{7ex}\\
\normalsize{Michihiro Naka
\footnote{email: naka@hep-th.phys.s.u-tokyo.ac.jp}
and Masatoshi Nozaki
\footnote{email: nozaki@hep-th.phys.s.u-tokyo.ac.jp}}
\vspace{5ex}\\
\normalsize{\it Department of Physics,\\
Faculty of Science, University of Tokyo,\\
Hongo 7-3-1, Bunkyo-ku, Tokyo 113-0033, Japan}
\end{center}
\vspace{5ex}
\abstract{We extend the construction of the boundary states
in Gepner models to the non-diagonal modular
invariant theories, and derive the same supersymmetric conditions 
as the diagonal theories.
We also investigate the relation between the microscopic charges
of the boundary states and Ramond charges of the B-type D-branes 
on the Calabi-Yau threefolds with one K\"ahler modulus in the large volume
limit.}

\end{titlepage}
\newpage

\section{Introduction}

\hspace*{4.5mm}
The theme of D-geometry \cite{d} is an interesting approach to 
D-branes on the curved space.
In this paper, we consider D-branes on the Calabi-Yau spaces in Type II
string theory by using the boundary states.
The BPS condition for the brane configurations on the Calabi-Yau spaces 
implies that the cycles on which D-branes wrap must be
the special Lagrangian submanifolds
for the middle-dimensional cycles and holomorphic 
for the even-dimensional cycles \cite{bbs}.
These cycles correspond to the A-type and 
the B-type boundary states \cite{ooy}.
It is well-known that Gepner model \cite{g} based on the 
$N=2$ minimal models describes the
exactly soluble string propagation on the Calabi-Yau 
space at the special symmetry enhanced point in the moduli space. 
The Gepner point is smoothly related to the geometric Calabi-Yau phase 
\cite{w}.
Recknagel and Schomerus \cite{rs} first considered D-branes in Gepner models.
They used the 
Cardy's construction \cite{c} for the bosonic subalgebra of the $N=2$ 
SCA and directly applied this to Gepner models with
the diagonal modular invariants.
Subsequently, the relation to the $N=2$ black holes was discovered \cite{gs}.
On the other hand, the Cardy's construction for the $A$-$D$-$E$ 
modular invariant theories \cite{ciz} has been established in \cite{pss,bppz}.
We apply this formalism to the boundary states in Gepner models in section 2.
We find that the boundary states satisfy the same 
supersymmetric condition as the diagonal cases. 
We also comment on the $K3$ compactification. 
Our construction includes all the known Gepner models 
classified in \cite{fkss}.
Then for these states, we compute the Witten index in Ramond sector 
which is interpreted as the geometric intersection form encoding
the quantization condition \cite{df}.
We can use this in order to count the number of the massless modes 
for the given configuration.

Some interesting approach was made in \cite{bdlr} in order to relate
the boundary states (in particular, the B-type states)
to the brane configurations on the quintic Calabi-Yau 
in the large volume limit.
The monodromy matrices found in \cite{cogp} and 
the intersection form were used in order to determine
the charge of the boundary states which represents the geometric brane
configurations. 
The same procedure was performed for the two-parameter 
Calabi-Yau threefolds with elliptic or $K3$ fibrations \cite{dr,kllw}.
This approach might bring some insights on the study of D-branes 
on the Calabi-Yau spaces, thus in section 3 
we pursue this procedure for the one-parameter Calabi-Yau
threefolds by using the calculation in \cite{kt}.
We calculate the charge for all the boundary states and find that 
there are D0-branes in some models as opposed to the quintic. 
We also compute the number of the moduli at the Gepner point for these states.
We find that the value seems to be larger than
the degrees of freedom for the deformation of the bundles. 
It is known that there are in some sense the nonperturbative modes 
which exist only at the Gepner point. 
Thus this calculation would include some modes 
which are unknown in the four-dimensional low energy effective theories.

After we had obtained the results of the present paper, we received a
paper \cite{sch}
which has some overlaps with section 3.
\section{Boundary states in Gepner models}

\hspace*{1.5mm}
In this section, we construct the boundary states for all the Gepner
models \cite{fkss}. 
Then we calculate the intersection form as the Witten index in Ramond sector.

\subsection{Boundary states}

\hspace*{4.5mm}
We begin with the rational CFTs with the chiral algebra 
${\cal A}$ $(={\cal A} _L={\cal A}_R)$ which have a finite set ${\cal
I}$ of the irreducible highest weight 
representation ${\cal H}_j$, for $j\in {\cal I}$. 
The Hilbert space of the bulk CFT is 
${\cal H}= \oplus _{(j,\bar j)\in {\rm Spec}}{\cal H}_j\oplus\bar{\cal
H}_{\bar j}$ with the multiplicity $N_{j\bar j}$ of the left and right
copies of ${\cal A}$.
Associated with this chiral algebra, 
there exist the chiral fields $W(z)$, $\bar W(\bar z)$ with the spin $s_W$.
In general, if there is an automorphism $\Omega$ 
that preserves the equal time commutators of the algebra, 
the boundary conditions can be written as
$W(z) = \Omega (\bar W)( \bar z)|_{z=\bar z}$.
The action of
$\Omega$ relates an irreducible representation ${\cal H}_{j}$ 
to another
irreducible representation ${\cal H}_{\omega(j)}$.
These boundary conditions are equivalent to those of the boundary states
satisfying 
$(W_n-(-1)^{s_W}\Omega(\bar{W}_{-n}))|\alpha \rangle_{\Omega}=0$,
where $W_n$ and $W_{-n}$ are the generators of 
the left- and right-moving chiral algebra ${\cal A}$.
In particular, $(L_n - \bar L_{-n})|\alpha\rangle =0$. 
These conditions
were solved by Ishibashi \cite{i} for the rational CFT.
The explicit solution $|j\rangle\rangle$ for the irreducible representation
${\cal H}_j$ of 
the algebra is given by
$|j\rangle\rangle
= \sum_N |j,N\rangle \otimes U \Omega |\widetilde{j,N}\rangle$
where the sum is over all the descendants of ${\cal H}_j$
and $U$ is an anti-unitary operator which acts only on the right-moving
generators as 
$U\bar{W}_nU^{-1}=(-1)^{h_W}\bar{W}_n$. 
There may be some multiplicity $N_{j j}$ for each Ishibashi states,
but we omit it in order to avoid the notational complications. 
This boundary state $|j\rangle\rangle$ couples to the Hilbert space
${\cal H}_j\otimes {\cal H}_{\omega(j)}$.
Thus it is labeled by ${\cal E}
=\{j|(j,\bar j =\omega(j))\in {\rm Spec} \}$, where ${\cal E}$ is called
the set of the exponents of the theory.

The Cardy's construction \cite{c} of the boundary states for the
non-diagonal modular invariant $SU(2)$ Wess-Zumino-Witten models has been
discussed in \cite{pss,bppz} and we explain it in the following
(we adopt the notation of \cite{bppz}).
Let $G$ be a Dynkin diagram of the $A$-$D$-$E$ type with the 
Coxeter number $g$ 
which is related to the level $k$ by $g=k+2$.
To each diagram, we associate an $n \times n$ adjacency matrix
$G_{\alpha \beta}$
($G_{\alpha \beta}=$ \# of the links between the node $\alpha$ and $\beta$),
where $n$ is the number of the nodes of its Dynkin diagram. 
\begin{equation}
\begin{array}{ccc}
{\rm Diagram}\;G & g=k+2 & j\in {\rm Exp}(G)\\
A_{n} & n+1 & 1,2,3,\ldots,n \\
D_{2n+1} &  4n &1,3,5,\ldots,4n-1,2n\\
D_{2n} & 4n-2 & 1,3,5,\ldots,4n-3,2n-1\\
E_6 & 12 &1,4,5,7,8,11\\
E_7 & 18 & 1,5,7,9,11,13,17\\
E_8 & 30 & 1,7,11,13,17,19,23,29\label{table1}
\end{array}
\end{equation}
This matrix is diagonalized in the orthonormal basis $\psi_{\alpha}^j $
which is labeled by the node $\alpha$ and the exponent $j$.
The table (\ref{table1}) is the list of the Coxeter number and the
exponents to each diagram.
For the diagonal cases, Cardy found the consistent boundary states
for the rational CFTs \cite{c}
with the help of the Verlinde formula
$N_{ij}^{k}=\sum_{\ell \in {\cal I}}{S_{i\ell} S_{j\ell}
S_{k\ell}^{*} \over S_{1\ell}}$. 
He required that the partition function on the strip which is periodic
in the time direction (topologically an annulus) with the boundary
conditions $\alpha $ and $\beta$ is equivalent, under the modular
transformation $\tau \to -1/\tau$, to that on the cylinder
between the boundary states $|\alpha\rangle $ and $|\beta\rangle$.
For the non-diagonal cases, the boundary states and their conjugate states are
given by 
\begin{equation}
|\alpha \rangle =\sum_{j\in {\rm Exp}(G)}\frac{\psi_{\alpha}^{j}}
{\sqrt{S_{1 j}}}|j\rangle\rangle ,\;\;\langle \beta 
|=\sum_{j\in{\rm Exp}(G)}\langle\langle j |
\frac{(\psi_{\beta}^{j})^*}{\sqrt{S_{1 j}}},
\end{equation}
where the set $\{\alpha\}$ and $\{\beta\}$ label the boundary states
which denote the nodes of a Dynkin diagram $G$.
The inner products of the Ishibashi states are defined by
$\langle\langle j'|\widetilde q ^{{1\over 2}{(
L_0+\bar L _0-{c \over 12})}}|
j\rangle\rangle =
\delta_{jj'}\chi_j(\widetilde q)$,
where $\widetilde q = e^{-2\pi i/\tau}$ and $\chi_j$ is the character
$\chi_j(\widetilde q) = 
{\rm Tr}_{{\cal H}_{j}}\widetilde q^{L_0-{c\over 24}}$. Then the 
partition function on the cylinder becomes 
$Z_{\alpha \beta}=\langle\beta| q ^{{1\over 2}({L_0+\bar L _0-{c \over
12}})}|
\alpha\rangle= 
\sum_{j\in {\rm Exp}(G)}\psi_{\alpha}^j
(\psi_{\beta}^j)^*\frac{\chi_j(\widetilde q)}
{S_{1j}}$.
On the other hand, the partition function as a periodic time evolution
on the strip with the boundary conditions $\alpha$ and $\beta$ becomes
$Z_{\alpha \beta}=\sum_{i\in {\cal I}}n_{\alpha \beta}^i \chi_i(q)$,
where $q=e^{2\pi i\tau}$ and $1\leq i \leq g$.
Under the S-transformation $\chi_j(\widetilde q)= \sum_{i\in {\cal I}} S_{ji}
\chi_i(q)$ (and $S=S^t$), we obtain the Cardy's equation
\begin{equation}\label{verlinde}
n_{\alpha \beta}^i =\sum_{j\in {\rm Exp}(G)}\frac{S_{ij}}{S_{1j}}
\psi_{\alpha}^j(\psi_{\beta}^j)^*.
\end{equation}
The fused adjacency matrices which are defined by 
$(V_i)_{\alpha}^{\;\;\beta}= n_{\alpha \beta}^i $
satisfy the fusion algebra
$V_i V_j =\sum_{k\in {\cal I}} N_{ij}^{k} \,V_k$ with the fusion
coefficients $N_{ij}^{k}$.
This algebra determines the
numerical values of $ n_{\alpha \beta}^i $ recursively by
$V_1=I$, $V_2=G$, $V_i =V_2V_{i-1}-V_{i-2}$ for
$3\leq i \leq g$
and we can see $V_g=0$.

Next we turn to Gepner models \cite{g}. 
Gepner used the so-called ``$\beta$ method'' 
to construct the supersymmetric Type II string compactifications
to $d+2$ dimensions by using the
tensor product of the $r$ $N=2$ minimal models at the levels $k_j \;
(j=1,\dots,r)$
with the internal central charge
$c=12-\frac{3}{2} d=\sum_{j=1}^{r}\frac{3k_j}{k_j+2}$. 
In each $N=2$ minimal model,  
the primary fields are labeled by the three integers $(\ell, m, s)$.
The standard range of $(\ell,m,s)$ is
$0\leq \ell \leq k$, $|m-s|\leq \ell$, $s\in\{-1,0,1\}$ 
and $\ell+m+s \in 2{\bf Z}$.
The representations with $s=0$ belong to the NS sector,
while those with $s=\pm 1$ to the Ramond sector.
The conformal dimension $h$ and the charge $q$ of the primary fields 
$\Phi_{m,s}^{\ell}$ are given by 
$h^{\ell}_{m,s} =\frac{\ell(\ell+2)-m^2}{4(k+2)} + \frac{s^2}{8}\,
({\rm mod}\,1),\;
q^{\ell}_{m,s} = \frac{m}{k+2} - \frac{s}{2} \,({\rm mod}\,2)$.
The chiral primary state and the anti-chiral primary state 
are labeled by
$(\ell,\pm \ell,0)$ in the NS sector and 
related to the Ramond ground states 
$(\ell,\pm \ell,\pm 1)$ by the spectral flow.
These minimal models
are invariant under the ${\bf Z}_n \times {\bf Z}_2$
discrete symmetry group with $n=k+2$ for $A_n$, $D_{2n+1}$, $E_6$ 
and $n=(k+2)/2$ for $D_{2n}$, $E_7$, $E_8$ \cite{fkss}.  
Their actions are
$g \Phi_{m,s}^{\ell}=e^{2\pi i \frac{m}{n}}\Phi_{m,s}^{\ell}$, 
$h \Phi_{m,s}^{\ell}=e^{2\pi i\frac{s}{2}}\Phi_{m,s}^{\ell}$. 
This ${\bf Z}_2$ symmetry is essentially the charge conjugation.
Then, let us introduce the vectors
$\lambda = (\ell_1,\ldots,\ell_r)\quad {\rm and}\quad 
\mu = (s_0; m_1,\ldots,m_r;s_1,\ldots,s_r)$
and the $2r+1$ dimensional vectors
$\beta_0=(1;1,\dots,1;1,\dots,1)$, 
and $\beta_j=(2;0,\dots,0;0,\dots,0,2,0,\dots,0)$ for $j=1,\dots,r$
which has 2 in the first and $(r+1+j)$-th entries 
and is zero everywhere else.
We use $s_0\in\{-1,0,1,2\}$
in order to characterize the irreducible ($s,o,c,v$) representations
of the $SO(d)$ current algebra generated by the $d$ external fermions. 
Then, define the inner product among the $2r+1$ dimensional vectors as
$\mu \bullet \widetilde{\mu}=
-\frac{d}{8}s_0\widetilde{s_0}-\sum_{j=1}^r\frac{s_j\widetilde{s_j}}{4}
+\sum_{j=1}^r\frac{m_j\widetilde{m_j}}{2(k_j+2)}$.  
The $\beta$-constraints are realized as
$q_{\rm tot}=2\beta_0\bullet \mu \in 2{\bf Z}+1,
\beta_j \bullet\mu \in {\bf Z}$. 
These are consistent only for $d=2,6$. For $d=4$, we have to replace $d$
with $d+2$ in the inner product in order to impose the 
consistent conditions. 
Note that the massless fields satisfy $2 \beta_0 \bullet  \mu = \pm 1$.
Using all the ingredients, we obtain the partition function
in Gepner models describing a
superstring compactified to $d+2$ dimensions as 
follows:
\begin{equation}\label{cpf}
Z=
\frac{1}{2}\frac{({\rm Im \tau})^{-d}}{|\eta (q)|^{2d}}
\sum_{b_0=0}^{2K-1}\sum_{b_1,\dots,b_r=0,1}
\sum_{\lambda,\mu}^{{\rm ev},\beta}
(-1)^{b_0}
M_{\lambda\lambda'}
\chi_{\mu}^{\lambda}(q)
\chi_{\mu+b_0\beta_0+\sum_{j=1}^r b_j\beta_j}^{\lambda'}(\bar{q})\;,
\end{equation}
where $K={\rm lcm}(2,k_j+2)$ and the character is
$\chi_{\mu}^{\lambda}(q)=\chi_{s_0}(q)\chi_{m_1s_1}^{\ell_1}(q)\dots
\chi_{m_rs_r}^{\ell_r}(q)$.
We denote by $M_{\lambda\lambda'}$ 
the products of the Cappelli-Itzykson-Zuber matrices \cite{ciz}.
The summation $\sum^{{\rm ev},\beta}$ is taken that the indices
satisfy the $\beta$-constraints and $\ell_j+m_j+s_j\in 2{\bf Z}$. 
We can check that eq.(\ref{cpf}) is invariant under the S-transformation:
\begin{eqnarray}\label{5}
S^{\,{\rm f}}_{s_0,s'_0} &=& {1\over2}\; e^{- i\pi {d\over2}
{s_0s'_0\over2}}\;,\\
S^{\,k}_{(\ell,m,s),(\ell',m',s')} &=& {1\over\sqrt{2} (k+2)} \;  
\sin \pi {(\ell+1)(\ell'+1)\over k+2} \,
e^{i\pi {mm'\over k+2}} \,e^{-i\pi {ss'\over2}}.
\end{eqnarray}
Note that eq.(\ref{5}) is correct only for $d=2,6$ and we have to replace
$d$ with $d+2$ in the case of $d=4$. 
For the $D_{2n}$, $E_7$, and $E_8$ case, the
field identification exist independently for the holomorphic and
the anti-holomorphic part, i.e.,
$\Phi_{m,s;\bar m ,\bar s}^{\ell;\bar \ell}=\Phi_{m+k+2,s+2;\bar m ,\bar
s}^{k-\ell;\bar \ell}$ and this amounts to 
the additional factor $1/2$ in front of the partition function \cite{fkss}. 
In the $D_{2n}$ case, this factor is canceled 
for the particular values $\ell =\bar \ell =k/2$. 
For all the cases, the discrete symmetry and the Hodge number for Gepner
models agree with those of the geometric Calabi-Yau manifolds \cite{fkss}.  

Then we construct the boundary states in Gepner models. 
In this case, the algebra ${\cal A}$ is generated by the $N=2$ SCA. 
There are the two boundary
conditions for the $U(1)$ current $J$ and the 
superconformal generators $G^{\pm}$
which preserve half of the spacetime supersymmetries \cite{ooy}.
They are called the A-type and the B-type boundary conditions.
The A-type boundary states corresponding to D-branes wrapping on 
the middle-dimensional cycles satisfy
\begin{equation}
(J_n-\bar{J}_{-n})|B\rangle=0,\quad
(G_r^++i\eta\bar{G}_{-r}^-)|B\rangle=0,\quad
(G_r^-+i\eta\bar{G}_{-r}^+)|B\rangle=0,
\end{equation}
and the B-type boundary states corresponding to D-branes wrapping on 
the even-dimensional cycles satisfy
\begin{equation}
(J_n+\bar{J}_{-n})|B\rangle=0,\quad
(G_r^++i\eta\bar{G}_{-r}^+)|B\rangle=0,\quad
(G_r^-+i\eta\bar{G}_{-r}^-)|B\rangle=0.
\end{equation}
The choice of $\eta=\pm 1$ corresponds to the choice of the spin structure. 
The A-type boundary states satisfy $q=\bar q$ and the B-type boundary
states satisfy $q=-\bar q$.
Due to the property of the anti-unitary operator $U$ and 
the mirror automorphism $\Omega$ 
\cite{gs}, the Ishibashi states $|j\rangle\rangle = \sum_N
|j,N\rangle \otimes U\widetilde{|j,N\rangle}$ satisfy the B-type boundary
conditions, while the states $|j\rangle\rangle = \sum_N
|j,N\rangle \otimes U\Omega\widetilde{|j,N\rangle}$ satisfy the
A-type boundary conditions, where the Ishibashi states are 
labeled by the quantum number for the $N=2$ primaries : $(\ell,m,s)$. 
One should note that they label the
irreducible representations of the bosonic sub-algebra of the $N=2$ algebras
rather than the full $N=2$ representations.
Before we describe the boundary states in Gepner models, 
it is convenient to change the notation of $\ell$ into
$ \ell+1 \in {\rm Exp}(G)$ in order to fit the range of $\ell$ to
start from zero.
Then let $|\lambda,\mu\rangle\rangle_{A}$
($|\lambda,\mu\rangle\rangle_{B}$)
be the tensor products of
$r$ A-type (B-type) Ishibashi states and the external part. 
Following the procedure of \cite{rs},
we write down the boundary states in Gepner models as
\begin{equation}
|\alpha\rangle\rangle_{\Omega} \equiv |S_0; (L_j,M_j,S_j)_{j=1}^r 
\rangle_{\Omega} 
= {1\over\kappa^{\Omega}_{\alpha}} 
\sum_{\lambda+1 \in {\rm Exp}(G), \mu}^\beta \delta_{\Omega} 
\;B^{ \lambda ,\mu}_{\alpha} \,
|\lambda ,\mu \rangle\!\rangle_{\Omega} ,\label{bbs}
\end{equation}
where $S_0, L_j, M_j, S_j$ are integers and
$\ell_j+1\in {\rm Exp}(G_j)$ for $1\leq j \leq r$, and the 
normalization constant $\kappa_{\alpha}^{\Omega}$ is determined later.
The explicit form of $B^{\lambda ,\mu}_{\alpha}$ is 
\begin{equation}\label{b}
B^{\lambda ,\mu}_{\alpha} = 
(-1)^{{s_0^2\over2}} e^{-i\pi {d\over2} {s_0S_0\over2} } 
\prod_{j=1}^r 
\sqrt{k_j+2 \over 2}{\psi_{L_j}^{\ell_j} \over
\sin^{{1\over2}} \pi {\ell_j+1\over k_j+2}} \,
e^{i\pi {m_jM_j\over k_j+2} } \,
e^{-i\pi {s_jS_j\over2} } \ .
\end{equation}
The symbol $\delta_{\Omega}$ ($\Omega = A,B$) denotes
the constraint which the Ishibashi state $|\lambda,\mu\rangle\rangle_\Omega$
must appear in the partition function (\ref{cpf}) \cite{rs}. 
For the A-type boundary states, this requires no constraint. For the
B-type boundary states, all the $m_j$ are the same modulo $k_j+2$.

We can evaluate the partition functions on the cylinder 
by using eq.(\ref{verlinde}) and the identification
$\chi_{m,s}^{\ell}(q)=\chi_{m+k+2,s+2}^{k-\ell}(q)$.  
The result for the A-type boundary states is
\begin{eqnarray}
Z_{\alpha \widetilde{\alpha}}^A(q)
&=&
\frac{1}{C_A}\sum_{\lambda'\in {\cal I},\mu'}^{\rm{ev}}
\sum_{\nu_0=0}^{2K-1}
\sum_{\nu_1,\dots,\nu_r=0,1}\:
(-1)^{S_0-\widetilde{S_0}+s_0'}\:
\delta_{S_0-\widetilde{S_0}+s_0'+\nu_0+2-2\sum_{j=1}^{r}\nu_j}^{(4)}
\nonumber\\
& \times &
\prod_{j=1}^{r}\:n_{L_j\widetilde{L_j}}^{\ell_j'} \:
\delta_{M_j-\widetilde{M_j}+m_j'+\nu_0}^{(2k_j+4)} \:
\delta_{S_j-\widetilde{S_j}+s_j'+\nu_0-2\nu_j}^{(4)}\: 
\chi_{\mu'}^{\lambda'}(q),\label{za}
\end{eqnarray}
and the result for the B-type boundary states is
\begin{eqnarray}
Z_{\alpha \widetilde{\alpha}}^B(q)
&=&
\frac{1}{C_B}\sum_{\lambda'\in {\cal I},\mu'}^{\rm{ev}}
\sum_{\nu_0=0}^{2K-1}
\sum_{\nu_1,\dots,\nu_r=0,1}\:
(-1)^{S_0-\widetilde{S_0}+s_0'}\:
\delta_{S_0-\widetilde{S_0}+s_0'+\nu_0+2-2\sum_{j=1}^{r}\nu_j}^{(4)}
\nonumber\\
&\times&
\delta_{{M-\widetilde M\over 2}+ \sum_{j=1}^r 
\frac{K'(m_j'+\nu_0)}{2k_j+4} }^{(K')}\:\prod_{j=1}^{r}
n_{L_j\widetilde{L_j}}^{\ell_j'} 
\delta_{M_j-\widetilde{M_j}+m_j'+\nu_0}^{(2)} 
\delta_{S_j-\widetilde{S_j}+s_j'+\nu_0-2\nu_j}^{(4)}
\chi_{\mu'}^{\lambda'}(q),\label{zb}
\end{eqnarray}
where ${\cal I}$ means 
that $\ell_j'$ 
runs from 0 to $k_j$; 
and $K'= {\rm lcm}(k_j+2)$, $M=\sum_{j=1}^r\frac{K'M_j}{k_j+2}$. 
The overall factors in eq. (\ref{za}), (\ref{zb}) are expressed as
\begin{equation}
{1\over C_A}=\frac{2^{\frac{r}{2}}\prod(k_j+2)}
{K\kappa_{\alpha}^A\kappa_{\widetilde{\alpha}}^A},\quad
{1\over C_B}=
\frac{2^{\frac{r}{2}}}{\kappa_{\alpha}^B\kappa_{\widetilde{\alpha}}^B},
\end{equation}
and we choose $\kappa_{\alpha}^{A}$ and $\kappa_{\alpha}^{B}$ 
to satisfy the Cardy's condition.
We find that for the non-diagonal cases, 
the supersymmetric conditions among two boundary states take the same form  
\begin{eqnarray}
S_0-\widetilde{S}_0& \equiv& S_j-\widetilde{S}_j\;\; ({\rm mod}\; 2)\;\;\;\;
{\rm for \;\; all}\;\;\;\; j=1,\;\dots,\,r\,,\\
Q(\alpha-\widetilde{\alpha})&=&
-\frac{d}{2}\frac{S_0-\widetilde{S}_0}{2}
-\sum_{j=1}^r\frac{S_j-\widetilde{S}_j}{2}
+\sum_{j=1}^r\frac{M_j-\widetilde{M}_j}{k_j+2} \in 2\bf{Z},\label{susy}
\end{eqnarray}
as the diagonal cases. 
This condition guarantees that there exists no tachyon in the open string
spectrum.

Next we discuss the $K3$ compactification
\footnote{In this case, the different method by using the spectral flow
invariant orbits was proposed in \cite{gjs}.}.
It turns out that the construction is straightforward.
The difference essentially lies on $d \in 4 {\bf Z}$.  
Some parts in the $\beta$-constraints and the S-modular transformation 
of the external part (\ref{5}) depend on $d$. 
Thus the boundary states in Gepner models for the $K3$ can be obtained 
only by replacing $d$ with $d+2$ and particularly substitute $d=4$ in the 
formula (\ref{b}), (\ref{susy}).
With all the things considered, we have constructed the boundary states
in all the Gepner models classified in \cite{fkss}
\footnote{There are some Gepner models in Type I string theory \cite{abpss}.
We thank A. Sagnotti for pointing out this. }. 

It is convenient to omit the external fermion index $s_0$ 
in order to consider only the internal CFT \cite{bdlr}. 
In this case, we adopt the ansatz for the 
modular invariant A-type and B-type boundary states with respect to
the internal part to have the same form as eq.(\ref{bbs}) but without the
external labels $s_0, S_0$, i.e.
$|\alpha\rangle\rangle_{\Omega}=|(L_j,M_j,S_j)_{j=1}^{r}\rangle_{\Omega}$.
Then we set the coefficients $B_{\alpha}^{\lambda,\mu}$ to be
\begin{equation}\label{b2}
B_{\alpha}^{\lambda,\mu}
=\prod_{j=1}^r
\frac{1}{\sqrt{2\sqrt{2}}}
\frac{\psi_{L_j}^{\ell_j}}{\sin^{1\over 2}\pi{\ell_j+1\over k_j+2}}
e^{i\pi\frac{m_jM_j}{k_j+2}}
e^{-i\pi\frac{s_jS_j}{2}},
\end{equation}
and we have to set
$\mu$, $\beta_0$ and $\beta_i$ to be the all $2r$-dimensional vectors
without the external indices 
(the inner product should be trivially modified). 
Also in this case, the $\beta$-constraints mean that 
the internal $U(1)$ charge is not necessarily odd-integer.
Then the construction is essentially the same.
We find that the condition that the two boundary states
$|\alpha\rangle\rangle$
and $|\widetilde{\alpha}\rangle\rangle$, with the same external part,
preserve the same supersymmetries is
\begin{equation}\label{ssc}
Q(\alpha-\widetilde{\alpha}):=
-\frac{S-\widetilde{S}}{2}+\frac{M-\widetilde{M}}{K'}
\in 2{\bf Z},
\end{equation}
where $S=\sum_{j=1}^r S_j$ and $M=\sum_{j=1}^r{K'M_j \over k_j+2}$. 

Before closing this subsection, we discuss the specific aspects about the 
diagonal cases to use in section 3
\footnote{Of course, we can make the case-by-case comments for the other cases.
}. 
First, $\psi_{L_j}^{\ell_j}$ in eqs.(\ref{b}),(\ref{b2}) is simply
the modular S-matrix.
From the Verlinde formula, $n^{\ell}_{L\widetilde L}$ in 
eqs.(\ref{za}),(\ref{zb}) becomes
$N^{\ell}_{L\widetilde L}$ 
which is the $SU(2)_k$ fusion rule coefficient : 
$N^{\ell}_{L\widetilde L}= 1\; {\rm for}\;
|L- {\widetilde L} |\le \ell\le\min\{L+{\widetilde L},
2k-L-{\widetilde L}\}\; {\rm and}\; \ell+L+{\widetilde L}\in 2{\bf Z}$ 
and otherwise 
$N^{\ell}_{L\widetilde L}=0$.
Then it follows from eqs.(\ref{za}),(\ref{zb}) that the integers
$(L_j,M_j,S_j)$ satisfy the condition 
$(\widetilde{L}_j-L_j)+
(\widetilde{M}_j-M_j)+(\widetilde{S}_j-S_j) \in 2\bf{Z}$
and the identification 
$(L_j,M_j,S_j) \sim (k_j-L_j,M_j+k_j+2,S_j+2)$.
Second, the boundary states have the
 ${\bf Z}_{k_j+2}$, ${\bf Z}_2$ symmetries 
$g_j : M_j \to M_j+2$, $h_j : S_j \to S_j +2$ 
inherited from the discrete symmetry in the $k_j$-th minimal models. 
Because of the $\beta$-constraints, the physically inequivalent choices for
$S_j$ are $S=0,2$ $({\rm mod}\; 4)$.
Thus the A-type boundary states are labeled by $|\{L_j\};\{M_j\};S\rangle$
and their discrete symmetries are $g_j$'s satisfying $\prod_{j=1}^rg_j=1$.
For the B-type boundary states, in addition to $S_j$,
the physically inequivalent choices for $M_j$ are also 
restricted and they are labeled by 
$M=\sum_{j=1}^r\frac{K'M_j}{k_j+2}$.
Then each discrete symmetry $g_j$ becomes 
$g_j=g^{{K'/(k_j+2)}}$ for $g\in {\bf Z}_{K'}$ and
the B-type boundary states are labeled by 
\begin{equation}
g^{M/2}h^{S/2}|\{L_i\}\rangle=|\{L_i\};M;S\rangle.
\end{equation}
We adopt the convention that
$(L_j,M_j,S_j)$ satisfy $\sum L_j+M+S\in 2{\bf Z}$.

\subsection{The intersection form}

\hspace*{4.5mm}
The CFT version of the intersection form in the classical geometry 
should be calculated as the Witten index $I_\Omega={\rm Tr}_R(-1)^F$ in 
the Ramond sector in the open string channel \cite{df}. 
When D-branes give rise to the particles in the macroscopic directions, 
this number is equivalent to 
the Dirac-Schwinger-Zwanziger symplectic inner product on their charges. 
This quantization condition was also proposed in \cite{gs}.

The Witten index corresponds to the amplitude between the RR 
boundary states with a $(-1)^{F_L}$ inserted in the closed string channel.
Only the Ramond ground states 
$\phi_{\ell +1,1}^{\ell} \simeq \phi_{-k+\ell-1,-1}^{k-\ell}$ 
contribute to the Witten index, 
thus we can evaluate it explicitly.
Due to the Verlinde formula and the S-matrices, 
we can extend the range of the superscript 
$\ell$ of $n_{L\widetilde{L}}^{\ell}$ so that $\ell$ 
has a period of $2k+4$.
Then we identify
$n_{L\widetilde{L}}^{-\ell-2}=-n_{L\widetilde{L}}^{\ell}$ and
set $n_{L\widetilde{L}}^{-1}=n_{L\widetilde{L}}^{k+1}=0$.
In the following, we present the results for $d=2,6$,
but we have only to interchange the conditions
on ${d \over 2} +r$ in order to obtain the results for $d=4$.
  
The result for the A-type boundary states is given by
\begin{equation}
I_A =
\frac{1}{\widetilde C_A}(-1)^{\frac{S-\widetilde{S}}{2}}\: 
\sum_{\nu_0=0}^{K-1}\:
(-1)^{({d \over 2}+r)\nu_0}
\prod_{j=1}^{r}\:
n_{L_j \widetilde{L}_j}^{2\nu_0+M_j-\widetilde{M}_j}.
\end{equation}
Then we consider the B-type boundary states.
When ${d\over 2}+r$ is even, the result is given by
\begin{equation}
I_B = \frac{1}{\widetilde C_B}(-1)^{\frac{S-\widetilde{S}}{2}}\:
\sum_{m_j'=0}^{2k_j+3}
\delta_{\frac{M-\widetilde{M}}{2}+
\sum\frac{K'(m_j'+1)}{2k_j+4}}^{(K')}
\prod_{j=1}^{r}\:
n_{L_j \widetilde{L}_j}^{-m_j'-1}\:,
\end{equation}
and when ${d\over 2}+r$ is odd, it becomes
\begin{eqnarray}
I_B &=& \frac{1}{\widetilde C_B}
(-1)^{\frac{S-\widetilde{S}}{2}}\:\sum_{m_j'=0}^{2k_j+3}\:
(-1)^{ {M-\widetilde M\over K'} +\sum
{(m_j'+1)\over k_j+2}}\nonumber\\
&\times&
{1\over 2}\:\delta_{{M-\widetilde M \over 2}+\sum
{K' (m_j'+1)\over 2k_j+4}}^{(K'/2)}
\prod_{j=1}^r\:
n_{L_j \widetilde{L}_j}^{-m_j'-1}\:.\label{iodd}
\end{eqnarray}
We choose the normalization
${\widetilde C_A}={\kappa_{\alpha}^A
\kappa_{\widetilde \alpha}^A K}$ and
${\widetilde C_B}={\kappa_{\alpha}^B
\kappa_{\widetilde \alpha}^B \prod_{j=1}^r(k_j+2)/K}$
to satisfy the Cardy's condition on the partition functions. 

The ${\bf Z}_2$ action $S\rightarrow S+2$ changes
the orientation of D-branes. 
The results for B-type boundary states depend only on $M-\widetilde{M}$ 
expected from the discrete symmetry.

We can count the number of the moduli of these boundary states by 
using the intersection form. 
The procedure was explained in \cite {bdlr}.
We can deal with all the $A$-$D$-$E$ cases
but we deal only with the diagonal case of the
B-type boundary states for the later use.

In the diagonal case, $n_{L \widetilde L}^{\ell}$ 
becomes the $SU(2)_k$ fusion coefficient $N_{L \widetilde L}^{\ell}$ 
and the intersection form can be easily rewritten in terms of the
generator of the discrete symmetry.
This fusion coefficient is expressed as \cite{dr} 
\begin{equation}
{\bf n}_{L \widetilde L}= g^{{|L-\widetilde L| \over 2}}
+g^{{|L-\widetilde L| \over 2}+1}
+\dots+g^{{L+\widetilde L \over 2}}
-g^{-1-{|L-\widetilde L| \over 2}}-\dots
-g^{-1-{L+\widetilde L \over 2}},
\end{equation}
or we can write ${\bf n}_{L\widetilde L}=t_L{\bf n}_{00}t_{\widetilde L}
^t$ in terms of the linear transformation $t_L=t_{L}^t=
\sum_{l=-L/2}^{L/2}g^l$ and ${\bf n}_{00}=(1-g^{-1})$. 
When ${d\over 2}+r$ is even, 
the intersection form for the B-type boundary states is given by 
\begin{equation}\label{ib}
I_B=\prod_{j=1}^r{\bf n}_{L_j \widetilde L_j} .
\end{equation}
When ${d\over 2}+r$ is odd, we can not use the formula (\ref{ib}) naively 
because the period of $M$ is not $2K'$ but $K'$ from eq.(\ref{iodd}).
Thus we have to multiply the
additional factor ${1\over 2}(1-g^{K'/2})$,
which comes from the trivial (0-th) minimal model
in order to use the above expression \cite{kllw}. 
Then the number of the moduli is given by the diagonal part of 
\begin{equation}\label{nom}
{1\over2}\prod_{j=1}^r{\bf \widetilde n}_{L_j\widetilde L_j}P-v,
\end{equation}   
where ${\bf \widetilde n}_{L\widetilde L}=
|{\bf n}_{L\widetilde L}|$ and 
$P$ is the matrix which comes from the trivial factor, $P=1$ when 
${d\over 2}+r$ is even and  $P={1\over2}(1+g^{K'/2})$ when ${d\over 2}+r$
is odd. 
We denote by $v$ the number of the vacuum states.  When ${d\over 2}+r$ is odd,
$v=2^{\nu}$ where $\nu$ is the number of $L_j$'s equal to $k_j/2$. When 
${d\over 2}+r$ is even, $v=2^{\nu-1}$ for $\nu>0$ and $v=1$ for $\nu=0$.

\section{Geometric interpretation of the boundary states}

\hspace*{4.5mm}
The boundary states constructed in the previous section represent
D-branes in the stringy scale, therefore it is interesting to relate
them with D-brane configurations in the large volume limit \cite{bdlr}.
We consider this procedure for the B-type 
D-branes on the one-parameter Calabi-Yau threefolds $X$
corresponding to the $4^41, 6^4, 8^33$ Gepner models 
with the diagonal modular invariants which we call $k=6,8,10$,
respectively as in \cite{kt}.
They are given by 
\begin{eqnarray}
W_0^{k=6}\!\!\!\!&=&\!\!\!\!z_1^6+z_2^6+z_3^6+z_4^6+2z_5^3=0, \nonumber\\
W_0^{k=8}\!\!\!\!&=&\!\!\!\!z_1^8+z_2^8+z_3^8+z_4^8+4z_5^2=0, \\
W_0^{k=10}\!\!\!\!&=&\!\!\!\!
z_1^{10}+z_2^{10}+z_3^{10}+2z_4^5+5z_5^2=0, \nonumber
\end{eqnarray}
in ${\bf WCP}_{(\nu_1,\dots,\nu_5)}$
where $\nu_i=k/n_i$ ($n_i$ appear in the form $z_i^{n_i}$)
and $h_{2,1}=103, 149, 145$ for $k=6,8,10$, respectively. 
The K\"ahler moduli space and the prepotential were 
considered in \cite{kt} following \cite{cogp} which computed 
the periods of the 3-cycles on the mirror $\hat{X}$
and the mirror map.
In order to obtain the mirror $\hat{X}$, consider the polynomial
\begin{equation}
W=W_0 - k \psi z_1z_2z_3z_4z_5 = 0,
\end{equation}
and then orbifoldize the resulting algebraic hypersurface by 
$G={\bf Z}_3 \times {\bf Z}_6^2, {\bf Z}_8^2 \times {\bf Z}_2, {\bf
Z}_{10}^2$ for $k=6, 8, 10$, respectively \cite{gp}.
The complex structure moduli space of $\hat{X}$ is a Riemann sphere 
with the three singularities and parameterized by $\psi^k$.
Basically, the BPS states wrapped on 3-cycles $Q_i[\Sigma_i]$ in $\hat{X}$
are characterized by the central charge $Z=Q_i \Pi_i$
where the periods are given by $\Pi_i=\int_{\Sigma_i}\Omega$
($\Omega$ is the holomorphic three-form in $\hat{X}$).
Each singularity in the moduli space gives
the monodromy which characterizes the period.

The procedure in \cite{bdlr} consists of the following steps:

(i)  Interpret the boundary states at the Gepner point as the generic
BPS states in the moduli space.

(ii) Use the monodromy matrices in order 
to relate the Gepner point and the large volume limit.

(iii) Translate the BPS charge vectors into the microscopic topological
charges using the Chern-Simons couplings.

This procedure involves the analytic continuation between the two
distinct regions in the moduli space, therefore the spectrum of the BPS states 
would be affected by the jumping and the marginal stability phenomena.
But we are not able to treat the dynamical aspects 
such as the stability and the existence of bound states here.
The precise form of this comparison depends on the choice of 
the path in the K\"ahler moduli space which goes 
from the Gepner point to the large volume limit
through the complex moduli space of the mirror. 
This is characterized by the flat 
$Sp(4, {\bf Z})$ connection provided by the special geometry.

We begin with the step (ii).
At the origin ($\psi^k=0$),
the modes has an additional ${\bf Z}_k$ global symmetry. 
This is an orbifold singularity in the moduli space.
The Gepner point corresponds to
this point in the K\"ahler moduli space of $X$. 
Let $A_G$ be the monodromy matrix induced by $\psi \to \alpha\psi$
around $\psi^k=0$ where $\alpha=e^{2\pi i/k}$.
This matrix satisfies $A_G^k=1$. 
If we choose a solution $\varpi_0(\psi)$ of Picard-Fuchs equations 
analytic near $\psi^k=0$,
the set of solutions $\varpi_i(\psi) = \varpi_0(\alpha^i\psi)$
will provide a basis of the period at the
Gepner point
\begin{equation}
\varpi=
-\frac{(2\pi i)^3}{{\rm Ord}G}
(\varpi_2,\varpi_1,\varpi_0,\varpi_{k-1})^t,
\end{equation}
where ${\rm Ord}G=3\cdot 6^2, 2\cdot 8^2, 10^2$
with the constraints 
$\varpi_0+\varpi_2+\varpi_4 =0,\varpi_1+\varpi_3+\varpi_5=0;\;
\varpi_i+ \varpi_{i+4}=0 (i=0,1,2,3);\;
\varpi_0+\varpi_2+\varpi_4+\varpi_6+\varpi_8=0
\footnote{There is a typo in \cite{kt}},
\varpi_i+\varpi_{i+5}=0  (i=0,1,2,3,4)$ for $k=6,8,10$, respectively.
Then let us consider the large volume limit 
which is mirror to the
large complex structure limit ($\psi^k \to \infty$).
There we obtain
$(\gamma\psi)^{-k}\simeq e^{2\pi i (B +i J)}$
where $\gamma =k\prod_{i=1}^5\nu_i^{-\nu_i/k}$, and
$B$ is the NS $B$-field flux around the 2-cycle forming a basis of
$H_2(X)$; and $J$ is the size of that 2-cycle.
The large volume basis is determined by the asymptotics
$\psi^k\to\infty$ as in \cite{bdlr},
\begin{equation}
\amalg=
\left(
\begin{array}{c}
\amalg_6\\
\amalg_4\\
\amalg_2\\
\amalg_0
\end{array}
\right)=
\left(
\begin{array}{c}
{\cal F}_2\\
{\cal F} _1\\
w_1\\
w_2
\end{array}
\right)
\simeq
w_2
\left(
\begin{array}{c}
-\frac{\kappa}{6}(B+iJ)^3\\
-\frac{\kappa}{2}(B+iJ)^2\\
B+iJ\\
1
\end{array}
\right),
\end{equation}
where $\kappa=3, 2, 1$ for $k=6, 8, 10$, respectively.
The coefficients give the classical volumes of the even-dimensional 
cycles.
The central charge corresponding to an integral vector 
$Q=(Q_6, Q_4, Q_2, Q_0)$ is
\begin{equation}
Z=Q_6\amalg_6 + Q_4\amalg_4 + Q_2\amalg_2 + Q_0\amalg_0.
\end{equation}
We can relate $\varpi$ to $\amalg$, i.e. 
$Z=Q\cdot\amalg=(Q_GM^{-1})\cdot(M\varpi)$.
This change of the basis by $M$ is $\amalg=M\varpi,\; Q=Q_GM^{-1}$.
The ${\bf Z}_k$ monodromy matrix in the large volume limit
is given by $A=MA_GM^{-1}$.
We obtain $M$'s as follows
\begin{equation}
M= 
\left(
\begin{array}{cccc}
0 & 1 & -1 & 0 \\
1 & 0 & -3 & -2 \\
-\frac{1}{3} & -\frac{1}{3} & \frac{1}{3}  & \frac{1}{3}\\
0  & 0 & -1 & 0
\end{array}
 \right),
\left(
\begin{array}{cccc}
0 & 1 & -1 & 0 \\
1 & 0 & -3 & -2 \\
-\frac{1}{2} & -\frac{1}{2} & \frac{1}{2}  & \frac{1}{2}\\
0  & 0 & -1 & 0
\end{array}
 \right),
\left(
\begin{array}{cccc}
0 & 1 & -1 & 0 \\
0 & -1 & -1 & -1 \\
-1 & 0 & 0 & 1\\
0  & 0 & -1 & 0
\end{array}
 \right),
\end{equation}
for $k=6,8,10$, respectively
\footnote{Of course, there is an undetermined $Sp(4,{\bf Z})$ ambiguity.
We can resolve this ambiguity by the following criteria\cite{bdlr}.
At the conifold point ($\psi^k=1$), 
the wrapped D3-brane becomes massless \cite{s}.
We choose the state which
is the mirror to the conifold point to be a D6-brane
with the trivial gauge bundle \cite{ps}.}.
We also obtain $A$'s as follows
\begin{equation}
A=\left(
\begin{array}{cccc}
-3 & -1 & -6 & 4 \\
-3 & 1 & 3 & 3 \\
1 & 0 & 1  & -1\\
-1  & 0 & 0 & 1
\end{array}
 \right),\;
\left(
\begin{array}{cccc}
-3 & -1 & -4 & 4 \\
-2 & 1 & 2 & 2 \\
1 & 0 & 1  & -1\\
-1  & 0 & 0 & 1
\end{array}
 \right),\;
\left(
\begin{array}{cccc}
-2 & -1 & -1 & 3 \\
0 & 1 & 1 & 0 \\
1 & 0 & 1  & -1\\
-1  & 0 & 0 & 1
\end{array}
 \right),
\end{equation}
for $k=6,8,10$, respectively.
Note that $A^4=-1$ for $k=8$ and $A^5=-1$ for $k=10$. 

Next we briefly comment on the step (iii).
The BPS charge lattice of the low energy effective theory is an integral 
symplectic lattice which can be identified with the middle cohomology
lattice of the mirror manifold $H^3(\hat{X},{\bf Z})$.
On the other hand, in the large volume limit, the lattice of the microscopic 
D-brane charges is an integral quadratic lattice identified with the K
theory lattice $K(X)$.
We have to construct a map between the low energy charges 
$Q$ and the topological invariants of the K theory class $\eta$
by exploiting the exact form of the Chern-Simons couplings.
The effective charges of D-branes are
measured by the Mukai vector $q \in H^{\rm even}(X)$ 
given by $q= ch(\eta)\sqrt{Td(X)}$.
The central charge associated with this state is then \cite{dr}
\begin{equation}
Z(t) = \frac{t^3}{6}q^0 - \frac{t^2}{2}q^2 + t q^4 -q^6.
\end{equation}
The comparison of these central charges gives the relation between the 
low energy charges and the topological invariants of $\eta$.
It is hard to see 
the behavior of each cycles in the case of one-parameter hypersurfaces.
Thus we can not care about this mapping and have to content ourselves with
determining only $(Q_6,Q_4,Q_2,Q_0)$.

Finally, we come to the step (i) which is the core of the method.
In principle, the above central charges should be compared with 
those of the boundary states, but such a comparison seems to be difficult.
It is useful to compute the interaction between the two D-branes 
in the open string channel.
This is canonically normalized because it is a partition function.
Therefore we consider the intersection form. 
We first express the known 
intersection form in the large volume limit
in terms of a natural basis at the Gepner point
which has the ${\bf Z}_k$ symmetry and 
then compare this with that of the boundary states.
Given the classical intersection form $\eta$ 
(of course, this is different from the above topological class)
in the large volume limit,
we can determine the intersection form in the Gepner basis,
$\eta_G=M^{-1}\eta(M^{-1})^t$
where $\eta$ is given by $\Sigma_6\cdot\Sigma_0=+1$ and
$\Sigma_4\cdot \Sigma_2=-1$ where $\Sigma_{2n}$ are 
the even-dimensional cycles in $X$. The intersection form
$\eta_G$ is not canonically normalized, but this does not concern us.
We can write down
the intersection form invariant under the ${\bf Z}_k$ symmetry. 
The ${\bf Z}_k$ generators act on the 3-cycles at the Gepner points
$\Sigma_i^G (i=0,\dots,k-1)$ as
$g\Sigma_i^G=\Sigma_{i+1}^G$. 
The results are 
\begin{eqnarray}
I_G^{k=6} \!\!\!\!&=&\!\!\!\! g(1+g)(1-g)^3,\nonumber\\
I_G^{k=8} \!\!\!\!&=&\!\!\!\! g(1+g)(1-g)^3(1+g^2),\\
I_G^{k=10} \!\!\!\!&=&\!\!\! g(1+g)(1-g)^2(1-g^5). \nonumber
\end{eqnarray}
On the other hand, we obtain 
the intersection form (\ref{ib}) among the $\sum L_i=0$ states as follows
\begin{eqnarray}
I_B^{k=6}\!\!\!\!&=&\!\!\!\!(1-g^4)(1-g^5)^4,\nonumber\\
I_B^{k=8}\!\!\!\!&=&\!\!\!\!(1-g^7)^4(1-g^4),\\
I_B^{k=10}\!\!\!\!&=&\!\!\!\!(1-g^9)^3(1-g^8)(1-g^5),\nonumber
\end{eqnarray}
where we have inserted the trivial factors in $k=8,10$ as explained before. 
We can identify the ${\bf Z}_k$ generator on the both basis to obtain 
\begin{equation}
I_B=(1-g)I_G(1-g)^t.
\end{equation}
>From this equation, we can relate the boundary states
$|\{0\};M;S\rangle$
to the basis of the periods $\varpi$ at the Gepner point.
Then the charges of the boundary states $Q_B$
is given by $Q_G=Q_B(1-g)$.

We summarize the procedure as follows.
First, we take $Q_G$ for $|\{0\};0;0\rangle$ to be $(0,1,-1,0)$ and
use the monodromy matrix in order to obtain the charge in the large volume
limit, $Q=Q_GM^{-1}$.
Then the state $|\{0\};0;0\rangle$ corresponds to the pure six-brane
$(1,0,0,0)$. 
We can obtain charges for different $M$ by acting $A^{-1}$ which
implements $g$ : $M \to M+2$.
The action $h$ : $S \to S+2$ is also implemented by $Q \to -Q$.
The charges of states with $L=\sum_iL_i>0$ can be obtained by
$Q_G=Q_B\prod_{j=1}^r t_{L_j}(1-g)$.
When $\sum L_i$ is odd, we also have to multiply the boundary states by
$g^{1/2}$.
By using the above procedure, we can determine the geometric charges for
all the boundary states.
We can also calculate the number of
supersymmetry preserving moduli
of the brane configuration
which consists of the same boundary states 
at the Gepner point by using eq.(\ref{nom}).
We find that it depends only on $\sum L_j$.  

It would be meaningless to array all the results, 
thus we give some interesting ones.
First, the values ($\Delta M=M-\widetilde M,\Delta S=S-\widetilde S$) 
which satisfy the supersymmetric 
condition (\ref{ssc}) are $(0,\,0)$ and $(k,\,2)$. 
Since all the $k$'s are even integers, the configuration between the state
with $\sum L_j=$ even and the state with $\sum L_j=$ odd
can not be supersymmetric
because of the condition $\Delta L +\Delta M+\Delta S \in 2{\bf Z}$. 
The conditions $A^4=-1$ for $k$=8 and $A^5=-1$ for $k$=10 
force the independent states to satisfy only $\Delta S =\Delta M = 0$. 
Thus the number of independent boundary states are $90,140,350$ for 
$k=6,8,10$, respectively.
But we found that some different boundary states have the same charge in the
large volume limit.
Then we give some results on
the geometric charges and the number of moduli of 
boundary states in the following tables.
On the left side, the boundary states 
$|\sum L_i\in 2{\bf Z};0;0\rangle$
which are supersymmetric with $|\{0\};0;0\rangle$ are included. 
On the right side, the boundary states $|\sum L_i\in 2{\bf Z}+1;1;0\rangle$
which are supersymmetric with 
 $|10000;1;0\rangle$ for $k=6$ 
or  $|1000;1;0\rangle$ for $k=8,10$ are included.

${\bf k=6}$
\begin{equation}
\begin{array}{cccccc}
L& Q_6 & Q_4 & Q_2 & Q_0 & {\rm moduli}\\
00000 & 1 & 0  & 0  & 0  & 0\\
20000 & -1 & 0 & -3 & 0 &  7\\
11000 & 0 & 0 &  -3 & 0 & 8\\
22000 & -2 & 0 & -6 & 0 & 22\\
11110 & -3 & 0 & -12 & 0 & 40\\
22200 & -4 & 0 & -12  &0 & 68\\
22110 & -6 & 0 & -18 & 0 & 94\\
22220 & -8 & 0 & -24 & 0 &208
\end{array}\quad
\begin{array}{cccccc}
L & Q_6 & Q_4 & Q_2 & Q_0 & {\rm moduli}\\
10000 & 2 & 1  & 2  & -4  & 1 \\
21000 & -2 & -1 & -9 & 1 & 15 \\
11100 & 0 & 0 &  -6 & -3 & 19 \\
22100 & -4 & -2 & -18 & 2 & 46\\
21110 & -6 & -3 & -27 & 3 & 63 \\
22210 & -8 & -4 & -36 & 4 & 140
\end{array}
\label{list}
\end{equation}

${\bf k=8}$
\begin{equation}
\begin{array}{cccccc}
L     & Q_6 & Q_4 & Q_2 & Q_0 & {\rm moduli}\\
0000 & 1 & 0  & 0  & 0  & 0\\
2000 & -1 & 0 & -2 & 0 & 6 \\
1100 & 0 & 0 &  -2 & 0 & 7 \\
3100 & -2 & 0 & -4 & 0 & 14 \\
\dots &  &  &  &  & \\
3300 & 0 & 0 & -4  &0 & 28 \\
\dots &  &  &  &  & \\
3333 & -8 & 0 & -32 & 0 & 496
\end{array}\quad
\begin{array}{cccccc}
L     & Q_6 & Q_4 & Q_2 & Q_0 & {\rm moduli}\\
1000 & 2 & 1  & 2  & -4  & 3\\
3000 & -4 & -2 & -8 & 6 & 6 \\
2100 & -2 & -1 &  -6 & 2 & 11\\
1110 & 0 & 0 & -4 & -2 & 15\\
\dots &  &  &  &  & \\
3310 & 0 & 0 & -8  & -4&60 \\
\dots &  &  &  &  & \\
3332 & 8 & 4 & 40 & 0 & 376
\end{array}
\end{equation}

${\bf k=10}$ 
\begin{equation}
\begin{array}{cccccc}
L     & Q_6 & Q_4 & Q_2 & Q_0 & {\rm moduli}\\
0000 & 1 & 0  & 0  & 0  & 0\\
2000 & 0 & 0 & -1 & 0 & 4 \\
1100 & 1 & 0 &  -1 & 0 & 5\\
0001 & -1 & 0 & -1 & 0 & 3\\
\dots &  &  &  &  & \\
4400 & 0 & 0 & -4  &0 & 36\\
\dots &  &  &  &  & \\
4441 & -8 & 0 & -24 & 0 & 392
\end{array}\quad
\begin{array}{cccccc}
L     & Q_6 & Q_4 & Q_2 & Q_0 & {\rm moduli}\\
1000 & 2 & 1  & 0  & -3  & 2\\
3000 & -2 & -1 & -2 & 2 & 6 \\
2100 & 0 & 0 &  -2 & -1 & 8\\
1110 & 2 & 1 & -2 & -4 & 12\\
\dots &  &  &  &  & \\
3111 & -2 & -1 & -8  &-1 & 50\\
\dots &  &  &  &  & \\
4331 & -8 & -4 & -24 & 0 & 316
\end{array}
\end{equation}

Only for $k=6$, 
there are the other supersymmetric states which are not listed in (\ref{list}).
On the left side, they are three independent states
$|00000;6;2\rangle$, $|11000;6;2\rangle$, $|11110;6;2\rangle$
with the charges $(-2,0,-3,0)$, $(-3,0,-6,0)$, $(-6,0,-15,0)$,
which are supersymmetric with $|00000;0;0\rangle$.
One should notice that the large number of the moduli such as
496 in $k=8$ may be related to the unknown degrees of freedom. 
The interpretation of this extra degrees of freedom seems to be 
the most urgent problem of this
approach.

We can also pick up some states whose charge have the only one component
except for the state $|00000;0;0\rangle$.
For $k=6$, there are no such states.
For $k=8$, $|3000;2;0\rangle$ has the charge $(0,0,0,-2)$. 
For $k=10$, $|4000;6;0\rangle,\;|4200;6;0\rangle,
\;|4440;6;0\rangle$ 
have the charges $(0,0,0,-2),\;(0,-2,0,0),\;(0,-8,0,0)$.
Thus we found D0-branes in $k=8, 10$ as opposed to quintic.
But this viewpoint may not be appropriate because we can make the 
linear combination of the boundary states to produce the various 
branes with the only one component.
Thus it would be important to consider the precise map between the two 
basis. 
We can also observe the phenomena that the non-BPS states 
in the large volume limit become the stable BPS states at the Gepner point. 
This may be related to the existence of the marginal stability lines
in \cite{d}.

In these settings, the marginal operators at the Gepner point 
would play the important role.
These boundary operator may have the superpotentials, with the
flat directions corresponding to the truly marginal operators.
In \cite{bdlr,kklm}, some first steps was made in order to 
compute the superpotentials on D-branes on Calabi-Yau. 
We hope to return this problem in the future work.

\section*{Acknowledgments} 

\hspace*{4.5mm}
We are grateful to T. Eguchi, Y. Matsuo for helpful discussions.
We also wish to thank J. Hashiba, M. Jinzenji, K. Sugiyama and Y. Satoh 
for discussions and encouragements.
The work of M. Naka is supported by JSPS
Research Fellowships for Young Scientists.

\end{document}